\documentclass[acmsmall]{acmart}

\AtBeginDocument{%
  }
\usepackage{enumitem}

\setcopyright{rightsretained}
\acmYear{2025} 
\acmDOI{10.1145/3744911}

\begin{document}

\title{Prompt-Hacking: The New p-Hacking?}

\author{Thomas Kosch}
\affiliation{%
  \institution{HU Berlin}
  \city{Berlin}
  \country{Germany}}
\email{thomas.kosch@hu-berlin.de}
\orcid{0000-0001-6300-9035}

\author{Sebastian Feger}
\affiliation{%
  \institution{TH Rosenheim}
  \city{Rosenheim}
  \country{Germany}}
\email{sebastian.feger@th-rosenheim.de}
\orcid{0000-0002-0287-0945}

\renewcommand{\shortauthors}{T. Kosch \& S. Feger}

\begin{abstract}
As Large Language Models (LLMs) become increasingly embedded in empirical research workflows, their use as analytical tools for quantitative or qualitative data raises pressing concerns for scientific integrity. This opinion paper draws a parallel between ``prompt-hacking'', the strategic tweaking of prompts to elicit desirable outputs from LLMs, and the well-documented practice of ``p-hacking'' in statistical analysis. We argue that the inherent biases, non-determinism, and opacity of LLMs make them unsuitable for data analysis tasks demanding rigor, impartiality, and reproducibility. We emphasize how researchers may inadvertently, or even deliberately, adjust prompts to confirm hypotheses while undermining research validity. We advocate for a critical view of using LLMs in research, transparent prompt documentation, and clear standards for when LLM use is appropriate. We discuss how LLMs can replace traditional analytical methods, whereas we recommend that LLMs should only be used with caution, oversight, and justification.
\end{abstract}






\maketitle

\section{Introduction}
Are Large Language Models (LLMs) helping or hurting research integrity? As their capabilities expand, the risks associated with their use in research become increasingly apparent. Rather than viewing LLMs as impartial or reliable tools, researchers must critically evaluate whether their use is appropriate. We argue that the inherent biases, variability, and susceptibility to manipulation make LLMs unsuitable for most data analysis tasks. This opinion parallels the risks of “prompt-hacking” to the practice of “p-hacking.” P-hacking is one of the most severe and widely recognized practices adversely affecting scientific integrity today. It provides a strong reference and foundation to stress the risks associated with questionable LLM practices and prompt-hacking within all computing disciplines and beyond. This serves as a basis for elaborating on whether we should trust LLMs as impartial data analysts. We say no and urge stricter usage standards when using LLM-based data analysis.

\section{Data Analysis in Empirical Research}
Empirical research in computer science relies on quantitative and qualitative methods to evaluate hypotheses from data collection. Quantitative studies often utilize statistical tools to validate results through numeric data, while qualitative studies, on the other hand, collect data from observations, interviews, and case studies to generate initial insights in a research field~\cite{https://doi.org/10.1111/ejed.12014}. Researchers state hypotheses or research questions before evaluating them through a study. After completing the data collection and analyzing the results, researchers contrast the results against their research questions and hypotheses, validating whether or not their results support the claims. Quantitative and qualitative research demands rigorous data collection, analysis, and interpretation practices to uphold the findings’ validity, reliability, and replicability. However, this careful process can be vulnerable to biases introduced through conscious or subconscious data manipulation techniques, whether by choice of variables, selective reporting, or biases inherent in analysis tools.

\section{Manipulating Research Results with p-Hacking}
In empirical research, p-hacking emerges as a substantial threat to scientific integrity. P-hacking occurs when researchers tune experimental data or statistical analysis to achieve a significant p-value, a statistical measure often used to confirm or reject hypotheses~\cite{doi:10.1177/0956797611417632}. Such tuning can involve selectively reporting variables, increasing sample sizes, or testing hypotheses that were changed after obtaining the results, which skew results towards significance, potentially misleading interpretations and conclusions. The consequences impact fields that rely on empirical evidence by eroding trust in findings that intensify the replication crisis, even resulting in documenting popular p-hacking strategies~\cite{doi:10.1098/rsos.220346}. As LLMs gain prominence as research analysis tools, the potential for similar manipulation through prompt-hacking grows. We are concerned that LLMs may not be trustworthy empirical data analysis tools.

\section{Prompt-Hacking: p-Hacking with LLMs}

LLMs are increasingly proposed as substitutes for traditional data analysis tools. However, their inherent biases, hallucinations, and variability make them fundamentally unreliable for tasks requiring impartiality and reproducibility~\cite{gibney2022ai}. Unlike statistical methods, which can be validated and replicated, LLM outputs depend highly on their training data and prompt phrasing, making them unsuitable for critical research processes. While their convenience may tempt researchers, we strongly caution against using LLMs for data analysis in most scenarios, as doing so risks compromising the validity and integrity of scientific findings. LLMs inherit biases and limitations from their training datasets~\cite{10.1145/3695765}, which can mislead interpretations and compromise research validity. While LLMs may appear to provide structured and reliable outputs, they are not designed to understand or evaluate the data context as a human researcher would. The risks include hallucinations, plausible but factually incorrect outputs, and reinforcement of entrenched cultural or institutional biases. Researchers must recognize that relying on LLMs for impartial analysis without critical oversight and validation could amplify errors and undermine scientific integrity. LLMs are not unbiased analysts but parrots whose output requires additional scrutiny.

We state that ``prompt-hacking'' closely resembles ``p-hacking,'' a problematic practice in data analysis where researchers tune variables, data, and statistical tests to achieve significant p-values. Prompt-hacking phenomena were introduced recently~\cite{10.1145/3695765, 10.1145/3673861}, and much like p-hacking, prompt-hacking may subconsciously encourage selective data manipulation. For example, researchers could keep modifying prompts to obtain outputs that support desired conclusions. Morris stated in a related opinion article, ``Prompting is a poor user interface for LLMs, which should be phased out as quickly as possible''~\cite{10.1145/3673861}. Researchers, especially non-LLM experts, may be unaware of how to prompt and how slight distinctions between prompting and natural-language interaction may create different research results. Unlike traditional research methods, LLM outputs vary dramatically depending on prompt phrasing and style. This variability in pseudo-natural language poses a challenge for reproducibility. Each prompt, even if only slightly altered, can yield different outputs, making it impossible to replicate findings reliably. As Morris noted, the lack of transparency in documenting prompt variations, validation processes, and final prompt selection biases can damage the scientific integrity of empirical studies. Prior studies explored using LLMs for data analysis~\cite{doi:10.1098/rsos.231053} and even for simulating human subject experiments~\cite{pmlr-v202-aher23a}. Yet, the prompting space is infinite, and subtle semantic or syntactic prompt changes may provide different research results. Morris highlighted that failing to report the number and history of unsuccessful prompts along with any distinguishing features of successful ones, neglecting to test whether slight prompt variations affect outcomes, and not verifying prompt consistency across different models, model versions, or repeated uses of the same model all represent significant oversights for research replicability when using LLMs.

Similarly, new concerns such as ``PARKing'' (Prompt Adjustments to Reach Known Outcomes) may arise, introducing additional risks to scientific integrity. In parallel to HARKing (Hypothesizing After Results Are Known)~\cite{10.1145/3173574.3173715}, we characterize PARKing as the practice of systematically modifying prompts until they yield results that align with pre-existing hypotheses, potentially creating a misleading picture of data that does not truly support the hypothesis. By encouraging prompt adjustments solely to support expected results, PARKing compromises the validity of outputs and hinders the credibility of findings.

\section{Are LLMs Appropriate for Data Analysis?}

While structured guidelines can mitigate some risks, researchers must be cautious of over-reliance on LLMs for tasks requiring impartiality. These models are different from human judgment and traditional qualitative or quantitative analysis. It is important to understand that even with improved transparency and documentation, the fundamental limitations of LLMs mean they should be used sparingly and primarily as a supplement to human analysis, not a substitute. LLMs are here to stay, and it is likely, or even already the case, that researchers rely on LLMs as a research tool~\cite{10.1145/3637436, doi:10.1098/rsos.231053}. We urge future research directions to advocate for careful LLM use. Although novel scientific insights can mitigate the risks of prompt-hacking, researchers must remain vigilant about the fundamental limitations of LLMs. Unlike p-hacking, where the misuse of statistical techniques can be uncovered through reproducible means, using LLMs as data analysts inherently introduces biases and inaccuracies, even when following guidelines, due to their non-deterministic output. We propose that researchers adopt a cautious mindset in cases where LLM use introduces unnecessary risks or could replace established rigorous methods. Only in limited and justified scenarios, where the benefits of LLMs outweigh their risks, should they be considered as tools for analysis.

\vspace{1em}
\noindent
\textbf{Evaluate the Necessity of LLMs:} Researchers should ask: Why are LLMs considered for this analysis? If traditional methods can achieve the same goals without introducing LLM-specific risks, LLMs should not be used.

\vspace{1em}
\noindent
\textbf{Assess Task Compatibility:} Determine if the analysis task aligns with LLM capabilities. LLMs are inappropriate for tasks requiring deep contextual understanding, impartial interpretation, or highly specialized domain knowledge.

\vspace{1em}
\noindent
\textbf{Standardizing Prompt Use in Data Generation and Analysis:} Clear guidelines should define to what extent LLMs are appropriate for data generation and analysis and when they are unsuitable. Establishing standards can mitigate inappropriate uses of LLMs in research. However, as LLMs evolve and update, these guidelines must be regularly reviewed and adapted to reflect changes in LLM capabilities and limitations.

\vspace{1em}
\noindent
\textbf{Review Ethical Implications:} Researchers must ensure that using LLMs does not compromise ethical standards, including avoiding cultural or systemic biases that may skew findings.

\vspace{1em}
\noindent
\textbf{Consider Reproducibility and Validity:} Reproducible, stable, and repeatable outputs are important data analysis components for ensuring reproducibility. To verify consistency, researchers should routinely repeat prompts and assess the stability of generated results over time. Researchers should also record the complete prompt creation process, including the steps, decisions, and the specific model used to develop the final prompting sequence. Any notable variations or required adjustments in prompt phrasing should be documented and reported transparently. This process allows researchers to account for potential fluctuations in LLM outputs, providing a clearer picture of their findings’ stability and reliability and enabling other researchers to reproduce and build on their work more accurately. LLMs should be avoided if LLM outputs are not consistently validated or replicated.

\vspace{1em}
\noindent
\textbf{Preregistration and Documentation of Prompts:} Based on the prompt stability process, researchers should preregister prompts and experimental protocols to ensure transparency. This includes documenting the sequence of prompts and their post-modifications deviating from the preregistration, helping prevent selective disclosure of prompts that favor specific hypotheses. While preregistration and documentation help reduce PARKing, the core issue is deciding whether LLMs should be used. Researchers must resist the temptation to repeatedly adjust prompts to align results with hypotheses. Instead, they should critically evaluate whether the task requires LLMs. The answer will often be that LLMs are unnecessary and potentially harmful. All these recommendations affect researchers, publication outlets, funding agencies, and research infrastructure providers. Infrastructure providers, including Zenodo and the Center for Open Science, must extend their features to capture preregistered prompts and metadata on target LLMs and their precise versions.

\section{Moving Towards Ethical and Reliable Use of LLMs in Research}
Whether to trust LLMs as impartial data analysts demands a clear and cautious stance: no, they should only be trusted with significant oversight. While their utility in accelerating specific research processes is undeniable, their inherent biases and variability show the need for a restrained approach and more research in this area. Researchers must prioritize the integrity of the scientific process above convenience, actively questioning the role and limitations of LLMs in empirical research. While the comparison to p-hacking highlights similarities in the risks of manipulation, it is essential to stress a key distinction: the outputs of LLMs are fundamentally shaped by their design and training, making them less objective than statistical tools. Unlike p-hacking, which often involves misused but inherently neutral techniques, prompt-hacking exploits tools that are not impartial by design. As such, even the ``correct'' use of LLMs in analysis cannot guarantee validity, demanding caution and critical oversight. The central question is not how to use LLMs responsibly but whether they should be used. The answer for most data analysis tasks is clear: avoid LLMs unless their use is essential and justifiable. The scientific community must resist the temptation to normalize LLM-based analysis and instead uphold the rigor and integrity of traditional methods.

\section{Acknowledgements}
This work is supported by the German Research Foundation (DFG), CRC 1404: “FONDA: Foundations of Workflows for Large-Scale Scientific Data Analysis” (Project-ID 414984028).

% \section{Author Information}
% Thomas Kosch is a Professor at HU Berlin, Berlin, Germany\newline
% Sebastian Feger is a Professor at TH Rosenheim, Rosenheim, Germany
    
\bibliographystyle{ACM-Reference-Format}
\bibliography{main}

\end{document}